\documentclass[aps,prb,twocolumn,groupedaddress,floatfix,showpacs]{revtex4}
\usepackage{graphics}
\usepackage{subfigure}
\usepackage{epsfig}
\usepackage{dcolumn}
\usepackage{bm}
\usepackage{overpic} 

\begin{document}

\title{Elemental distribution and oxygen deficiency of magnetron sputtered ITO films}

\author{Annett Th\o gersen}
\author{Margrethe Rein}
\author{Edouard Monakhov}
\author{Jeyanthinath Mayandi}

\affiliation{Institute for Energy Technology, Department of Solar Energy, Instituttveien 18, 2008 Kjeller, Norway}

\author{Spyros Diplas}
\affiliation{SINTEF Materials and Chemistry, P.B 124 Blindern, N-0314 Oslo, Norway, and Centre for Material Science and Nanotechnology, University of Oslo.}

\date{\today}

\begin{abstract}
The atomic structure and composition of non-interfacial ITO and ITO-Si interfaces were studied with Transmission Electron Microscopy (TEM) and X-ray Photoelectron Spectroscopy (XPS). The films were deposited by DC magnetron sputtering on mono-crystalline p-type (100) Si wafers. Both as deposited and heat treated films consisted of crystalline ITO. The ITO/Si interface showed a more complicated composition. A thin layer of SiO$_x$ was found at the ITO/Si interface together with In and Sn nanoclusters, as well as highly oxygen deficient regions, as observed by XPS. High energy electron exposure of this area crystallized the In nanoclusters and at the same time increased the SiO$_x$ interface layer thickness.

\end{abstract}

\maketitle

\section{Introduction}

Technological developments in photovoltaics such as the use of Transparent Conducting Oxides (TCO) like Indium Tin Oxide (ITO) demand sufficient characterization of thin films, surfaces, and interfaces. An In$_2$O$_3$ thin film has an optical band gap of more than 3.4 eV \cite{xirouchaki:intro, kin:intro}, with low resistivity \cite{ando:intro, chopra:intro}, high transparency (87\%)\cite{ando:intro} in the visible range, and compatibility with fine patterning processes \cite{otha:intro}. ITO can be used as electrodes for flat panel displays \cite{tahar:intro}, including Liquid Crystal Displays (LCD) \cite{otha:intro}, transparent electrodes for light-emitting diodes \cite{kin:intro}, thin film gas sensors, solar cells \cite{kobayashi:intro}, and as anodes in Organic Light Emitting Diodes (OLED). The structure and composition of ITO films has been the focus of research the last years \cite{Soonkim:in, kimho:intro, ishida:o, fan:ox}. The through thickness composition in various types of ITO films is an important issue, because of its impact on the device properties when used in photovoltaic applications. 

In modern microelectronic systems consisting of thin films deposited on substrates it is important to have a stable film/substrate interface. Interface reactions may occur during thin film deposition involving high energy incident species in the substrate surface or during subsequent high-temperature post deposition processing \cite{cleva:ITO}. The deposition of ITO on Si tends to create a thin layer of SiO$_x$ at the ITO/Si interface \cite{Quentin:siox}. This oxide is expected to change the properties of the interface, making it unsuitable for ohmic contact applications. However, it might be utilized for in-situ formation of ITO-metallized gate oxides \cite{cleva:ITO}. Kobayashi et al.\cite{intro:NC} reported formation of metallic In when they deposited ITO normal to the Si substrate. The formation of metallic In in ITO films changes their work function, which in turn may strongly affect the I-V characteristics of the ITO/Si solar cells \cite{intro:NC}. 

In this work we attempt to gain a better understanding of the composition of ITO films deposited by DC magnetron sputtering on Si as well as the structure of the ITO-Si interface, High Resolution TEM (HRTEM), Energy Filtered TEM (EFTEM), and XPS were used to study the composition of the non-interfacial ITO and the ITO/Si interface.

\section{Experimental}

ITO films were deposited on monocrystalline p-type Si (100) substrates by an industrially designed DC magnetron sputtering equipment made by Leybold Optics. Before deposition, the samples were cleaned in an ultra sound bath with DI water for 30 min and dried with compressed air. An additional HF dip for one minute and subsequent rinsing in DI-water for two minutes should provide an oxide free surface. The samples were mounted on the sputtering carrier and loaded into the load lock. The pressure in the load lock was 10$^{-6}$ mbar when the samples were heated at 100$^\circ$C for 30 min, and the carrier was transferred from the load lock to the process chamber. The base pressure in the process chamber was in the order of magnitude 10$^{-7}$ mbar, and the working pressure was always kept at 3.1$\cdot$10$^{-3}$ mbar. The flow of argon and oxygen was 180 sccm and 6 sccm (3.2\% of the total flow), respectively. The weight ratio of the indium-tin-oxide (In$_2$O$_3$/SnO$_2$) target was 90\% vs. 10\%. The ramp up time of the target was 15 min and the power density used was 5.3 W/cm$^2$. There was only one passing of the carrier in front of the target, and the carrier velocity was 0.7 m/min. The target to substrate distance was 80 mm. The sputtering experiments were performed without intentional substrate heating during deposition. A maximum temperature of $\sim$80$^\circ$C during deposition was measured by the use of Thermax thermostrips. After deposition, the samples were cut in two pieces, and one half of each sample was annealed on a hot plate at 300$^\circ$C for 15 min in ambient air.

\begin{figure}
  \begin{center}
    \includegraphics[width=0.4\textwidth]{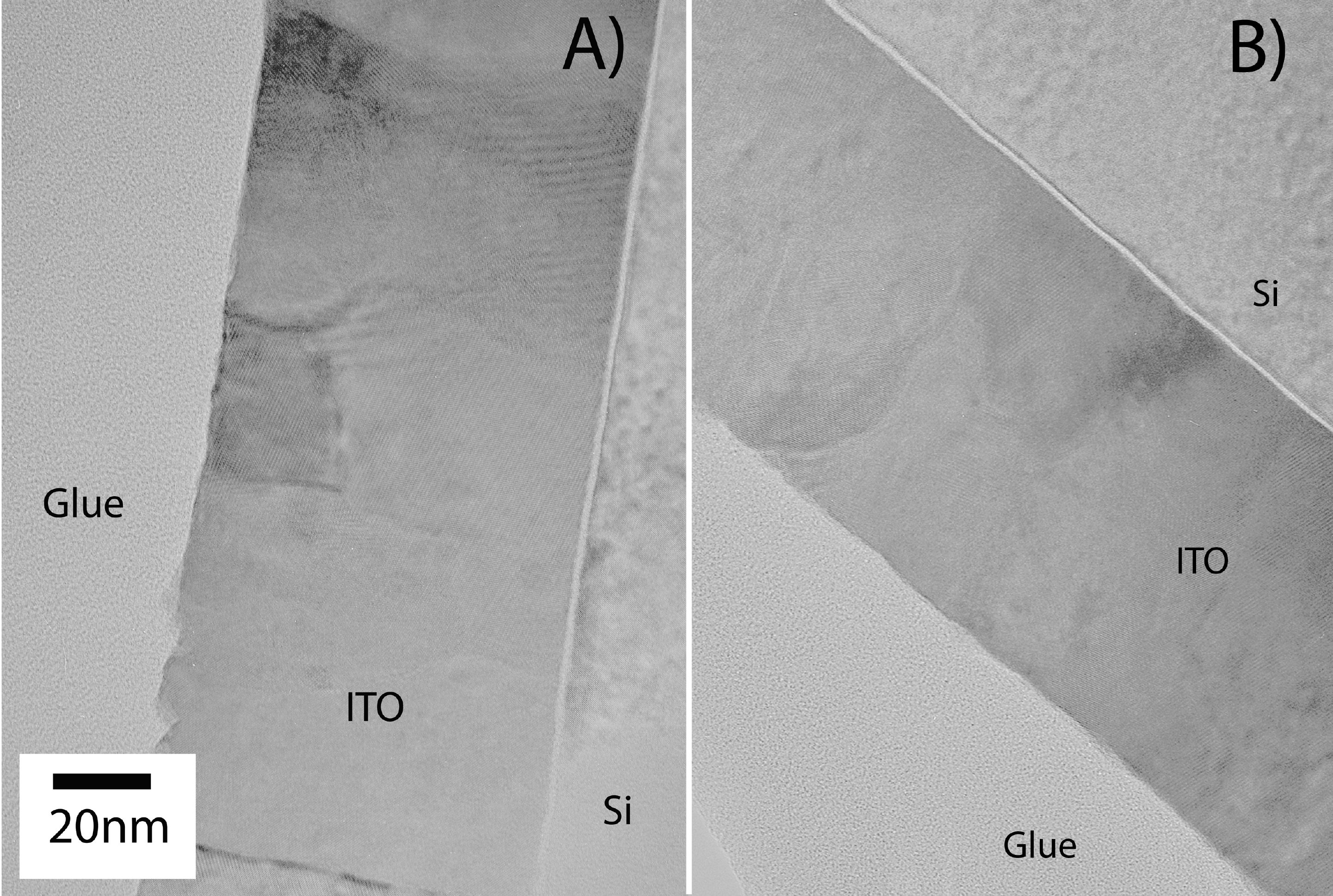}
   \caption{HRTEM image of the (A) as-deposited and (B) heat treated ITO layer on a Si substrate.}
    \label{figure:stor}
  \end{center}
\end{figure}

\begin{table}[hp]
\caption{Binding energy values for non-interfacial ITO, including the reference values for crystalline ITO, amorphous ITO, In(OH)$_x$, pure In, SnO$_2$, SnO, and pure Sn. }
\begin{center}
\noindent
\begin{tabular}{llll}
\hline
\hline
Composition\, & In-3d$_{5/2}$ (eV) \, & O-1s (eV) \, & Sn-3d$_{5/2}$ (eV) \, \\
\hline
In$_{II}$ & 444.9 & & \\
In$_{III}$ & 446.0& & \\
O$_{I}$ & & 530.5 & \\
O$_{II}$ & & 531.8 & \\
Sn$_{II}$ & & & 486.8 \\
Sn$_{III}$ & & & 488.0 \\
\hline
ITO-cryst. & 444.6\cite{zhua:in} \footnotemark[1] & 530.1\cite{xps:in} \footnotemark[2] & 486.4 \footnotemark[3]\\
ITO-a & 445.0\cite{xps:in} \footnotemark[1] & 531.4\cite{xps:in} \footnotemark[4] & \\
In(OH)$_x$ & 446.0\cite{xps:in} \footnotemark[1] & 532.6\cite{xps:in} \footnotemark[4] & \\
In & 443.7\cite{zhua:in} \footnotemark[1] & & \\
SnO$_2$ & & 530.6\cite{kimho:intro} \footnotemark[4] & 487.1\cite{xps:in} \footnotemark[5]\\
SnO & & 530.4\cite{Jimenez:sn} \footnotemark[4] & 486.2\cite{xps:in} \footnotemark[5]\\
Sn & & & 485.0\cite{Ke:sn} \footnotemark[5]\\
\hline
\hline
\end{tabular}
\end{center}
\footnotetext[1]{FWHM: 1.4 $\pm$ 0.1 eV \cite{xps:in}}
\footnotetext[2]{FWHM: 1.3 $\pm$ 0.1 eV \cite{xps:in}}
\footnotetext[3]{Reference \cite{oxdef:1}}
\footnotetext[4]{FWHM: 1.6 $\pm$ 0.1 eV \cite{xps:in}}
\footnotetext[5]{FWHM: 1.5 $\pm$ 0.1 eV \cite{xps:in}}
\label{table:ref}
\end{table}

Cross-sectional TEM samples were prepared by ion-milling using a Gatan precision ion polishing system with 5 kV gun voltage. The samples were analysed by HRTEM and EFTEM in a 200 keV JEOL 2010F microscope with a Gatan imaging filter and detector. The spherical (C$_s$) and chromatic aberration (C$_c$) coefficients of the objective lens were 0.5 mm and 1.1 mm, respectively. The point to point resolution was 0.194 nm at Scherzer focus ($-42$ nm). XPS was performed in a KRATOS AXIS ULTRA$^{DLD}$ using monochromatic Al K$_\alpha$ radiation ($h\nu=1486.6$ eV) on plan-view samples at zero angle of emission (vertical emission). The X-ray source was operated at 10 mA and 15 kV. The spectra were peak fitted using Casa XPS \cite{casa:xps} after subtraction of a Shirley type background. The Si-2p photoelectrons in ITO have a mean free path ($\lambda$) of 2.35 nm. This results in a photoelectron escape depth of 3$\lambda$ = 7.05 nm. Assuming uniform Ar$^+$ etching, an etching rate of about 0.04 nm/sec. was estimated by measuring the ITO thickness on cross sectional TEM samples, observing the Si-2p peak from XPS survey spectra as well as considering the photoelectron escape depth.

\section{Results and Discussion of non-interfacial ITO}
\label{section:bulkxps} 

\begin{table}[hp]
\caption{Atomic percentages and compositional ratios of both as deposited and heat treated non-interfacial ITO.}
\begin{center}
\noindent
\begin{tabular}{lll}
\hline
\hline
Element \, & as deposited \, & heat treated \, \\
\hline 
In$_ {II}$ & 34.0 & 33.4 \\
In$_ {III}$ & 5.0 & 4.8 \\
O$_{I}$ & 48.3 & 48.4 \\
O$_{II}$ & 10.2 & 10.9 \\
Sn$_{II}$ & 2.0 & 2.1 \\
Sn$_{III}$ & 0.5 & 0.6 \\
\hline
(O$_I$ + O$_{II}$)/ (In$_{total}$+Sn$_{total}$) & 1.41 & 1.45 \\
O$_I$/(In$_ {II}$ + Sn$_ {II}$) & 1.34 & 1.36 \\
O$_{II}$/(In$_ {III}$ + Sn$_ {III}$) & 1.83 & 2.02 \\
O$_{II}$/O$_{I}$ & 0.21 & 0.22 \\
\hline
\hline
\end{tabular}
\end{center}
\label{table:ratiobulk}
\end{table}

Figure \ref{figure:stor} shows TEM images of both the as deposited and heat treated sample. Both samples are crystalline, with a layer thickness of about 80 nm. Usually ITO is amorphous when sputtered at room temperature (RT) \cite{spyros:2}, but crystalline ITO has also previously been shown to form upon deposition at room temperature \cite{M:1, M:2, M:3, M:4}. Crystallization may be influenced by the gas pressure and the target to sample distance. A suitable gas pressure has to be chosen so as to minimize collisions between the sputtering gas and the sputtered target atoms, and the distance between target and the sample should not be too large in order to preserve the kinetic energy of the sputtered particles. By tuning these parameters it is feasible to obtain a sufficient surface mobility of the sputtered atoms and, hence, increase nucleation sites\cite{M:1, M:2, M:3}. It is also established that crystallization of DC sputtered ITO at room temperature (RT) depends on the thickness of the sputtered film. Moreover, it has been suggested that the substrate type affects the critical thickness of the film required for crystallization \cite{M:4}. It is established that the crystallization temperature of ITO is of the order of 150-160$^\circ$C \cite{ito:temp1, ito:temp2}, whilst in this work the measured sample temperature during deposition was 80$^\circ$C. 

Crystalline ITO may also form by plasma enhanced crystallization. In this context the presence of crystallinity in the as deposited samples could be explained as follows. The distance between target and substrate is 8 cm. At this distance, the plasma surrounding the target also covers parts of the wafer. It is therefore possible that the crystallization of the as deposited ITO layer is enhanced by the plasma.

\begin{figure}
  \begin{center}
    \includegraphics[width=0.5\textwidth]{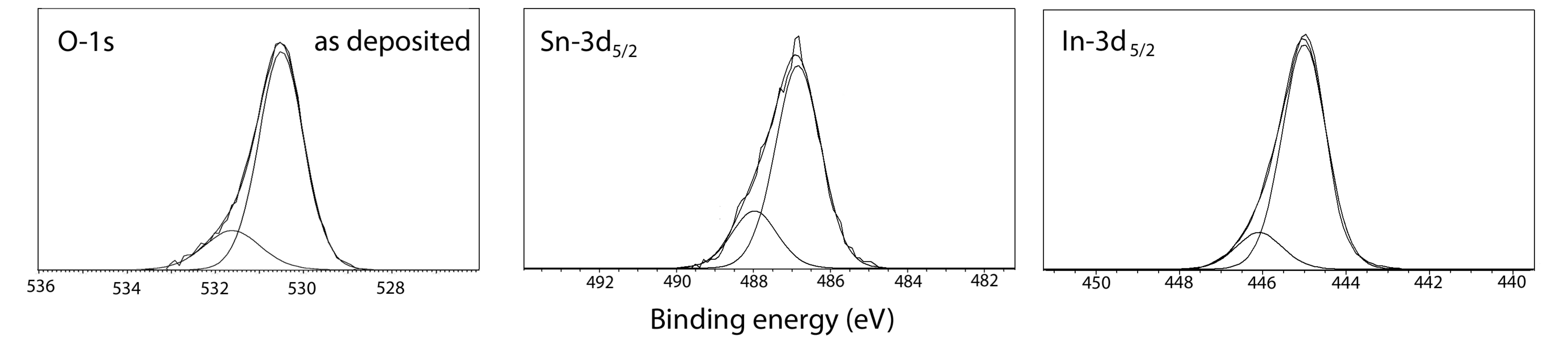}
   \caption{XPS spectra of A) O-1s, B) Sn-3d$_{5/2}$, and C) In-3d$_{5/2}$ of the as deposited bulk sample.}
    \label{figure:xpsbulk}
  \end{center}
\end{figure}

XPS was used to identify the composition and chemical state of the elements in the ITO. The XPS high resolution peaks of the O-1s, Sn-3d$_{5/2}$, and In-3d$_{5/2}$ for the two samples are shown in Figure \ref{figure:xpsbulk}. The O-1s spectra have been fitted with two peaks located at a binding energy of 530.5 eV and 531.8 eV. The Sn-3d$_{5/2}$ spectra have been fitted with two peaks at 486.8 eV and 488.0 eV, and the In-3d$_{5/2}$ at 444.9 eV and 446.0 eV; see Table \ref{table:ref}. The peaks have been labelled as to distinguish them during discussion. The origin of the Sn and In components labelled as In$_{II}$, Sn$_{II}$, In$_{III}$, Sn$_{III}$, and will be discussed in detail in this chapter. The In$_{I}$ and Sn$_{I}$ components will be discussed in Section \ref{section:itosi}. The peak positions and reference values are presented in Table \ref{table:ref}.

The binding energies of In$_{II}$ and In$_{III}$ are very close to the reference values for amorphous ITO and In(OH)$_x$, respectively. However, since TEM images revealed that most of the ITO was crystalline, the In$_{II}$ and Sn$_{II}$ peaks are most likely due to crystalline ITO. The two peaks have therefore been fitted with an asymmetry, due to the metallic character of ITO. The asymmetry of the peak was varied in order to see if the In$_{III}$ component could only be an expression of the asymmetry of the In$_{II}$ component. However, fitting the spectra with only one peak did not give satisfactory peak fitting. Kim et al.\cite{Soonkim:in} fitted the In-3d$_{5/2}$ peak with two components, located at 444.08 eV and 445.24 eV (energy separation 1.16 $\pm$ 0.1 eV), attributed to crystalline and amorphous ITO, respectively. The energy separation between the two fitted In-3d components in this work is 1.1 eV. This indicates that the small In-3d$_{5/2}$ peak at 446.0 eV may be attributed to amorphous ITO. The In$_{II}$/In$_{III}$ relative fractions are 90 at. \% and 10 at. \%, respectively. Since the Sn$_{II}$ peak corresponds to crystalline ITO, the Sn$_{III}$ may be attributed to Sn in amorphous ITO. The binding energies of the O$_{I}$ and O$_{II}$ peaks fit also well with that of crystalline and amorphous ITO, respectively. 

In order to fully determine the chemical state of the oxide, the electroneutrality principle was used \cite{kimho:intro}. The doubly-charged O$^{2-}$ and singly charged OH$^-$ will have different stoichiometry when combined with the lattice cation. Substitutional Sn in the In$_2$O$_3$ oxide will have an (O$_I$ + O$_{II}$)/(In + Sn) theoretical ratio of 1.5, while Sn interstitial will be 1.55 when the oxygen is in the form of O$^{2-}$ \cite{kimho:intro}. If the anions in the oxide are in the form of OH$^-$ instead of O$^{2-}$, the ratio will be 3.0 \cite{kimho:intro}. The atomic percentages of the different elements in ITO bulk as well as their calculated ratios for the as deposited (asd) and heat treated (ht) samples are presented in Table \ref{table:ratiobulk}. If O$_I$, In$_ {II}$, and Sn$_ {II}$ correspond to crystalline ITO (with the oxygen present as O$^{2-}$), the O$_I$/(In$_ {II}$ + Sn$_ {II}$) ratio would be about 1.5. The actual ratio for the as deposited sample was 1.34 and for the heat treated sample 1.36. The components that may correspond to amorphous ITO have a ratio of O$_{II}$/(In$_ {III}$ + Sn$_ {III}$) = 1.83 and 2.02 for as deposited and heat treated, respectively. This ratio is far higher than what is expected for amorphous ITO (about 1.5). When using the sum of all, the calculated ratios (O$_I$ + O$_{II}$)/(In$_ {II}$ + In$_ {III}$ + Sn$_ {II}$ + Sn$_ {III}$) 1.41 (asd) and 1.45 (ht), which are closer to the value for crystalline ITO. This suggest that the O$_I$, O$_{II}$, In$_ {II}$, In$_ {III}$, Sn$_ {II}$, and Sn$_ {III}$ all result from crystalline ITO, and the presence of In(OH)$_3$ is also excluded.

\begin{figure}
  \begin{center}
    \includegraphics[width=0.5\textwidth]{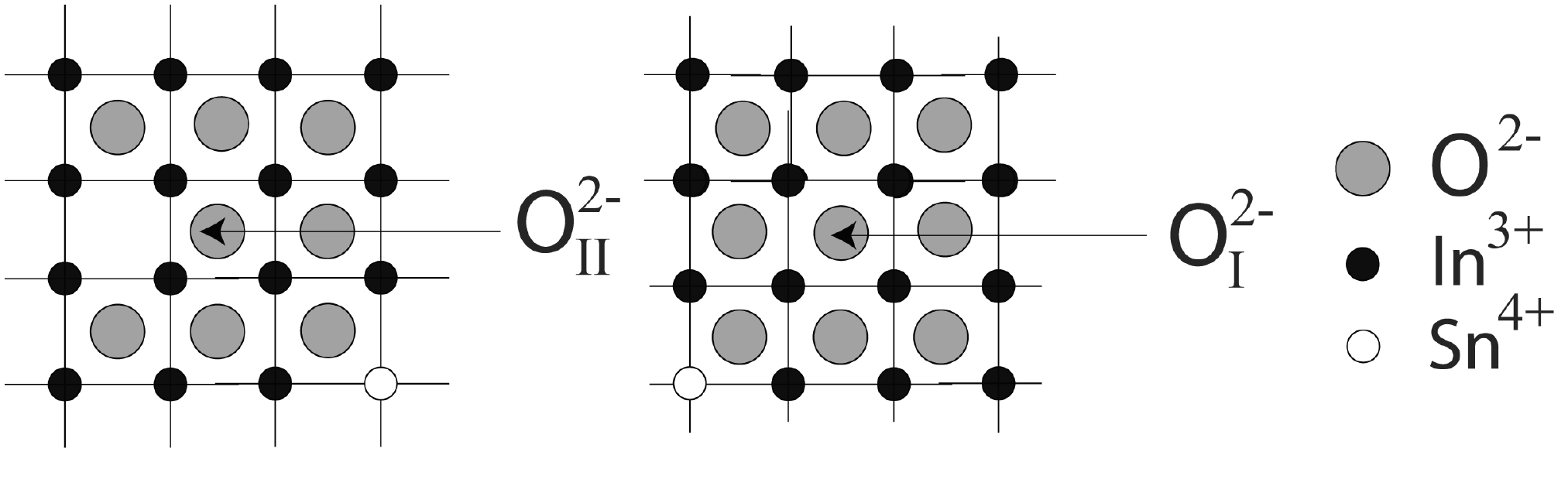}
    \caption{Two different kinds of O$^{2-}$ lattice ions located in different regions. The higher binding energy O$_ {II}^{2-}$ arises from the oxygen-deficient region.
region. Adapted from Kim et al. \cite{kimho:intro}.}
    \label{figure:ox}
  \end{center}
\end{figure}

Application of the electroneutrality principle contradicts our initial assignment of the In$_ {III}$, Sn$_ {III}$, and O$_ {II}$ compounds. However, crystalline ITO may have contributions from both O$_I$ and O$_{II}$ peaks. Kim el al. \cite{kimho:intro} showed that O$_I$ and O$_{II}$ may originate from two different types of O$^{2-}$ ions (O$^{2-}_I$ and O$^{2-}_ {II}$); see Figure \ref{figure:ox}. It has been suggested that a double O$^{2-}$ peak is common for oxides containing cations in multiple valence states \cite{kimho:intro, Hirokawa:o}. A higher binding energy of the O$_{II}^{2-}$ ions may be due to presence of oxygen in oxygen-deficient regions. This means that these oxygen ions do not have neighbouring In atoms with full complement of six nearest neighbour O$^{2-}$ ions \cite{kimho:intro, fan:ox}; see Figure \ref{figure:ox}. According to Fan et al.\cite{fan:ox}, the O$^{2-}_{II}$ peak has a about 1.5 eV higher binding energy than O$^{2-}_{I}$. The chemical shift between the O$_{I}$ and O$_{II}$ peaks in Figure \ref{figure:xpsbulk} is 1.3 $\pm$ 0.1 eV. The O$_{II}$ peak may therefore also contain contributions from oxygen deficient regions in the ITO. As a consequence, the O$_{II}$/O$_{I}$ ratio may be used to determine the oxygen deficiency of the material \cite{kimho:intro}. This ratio is 0.2 $\pm$ 0.1 for the as deposited and 0.2 $\pm$ 0.1 for the heat treated sample (see Table \ref{table:ratiobulk}). The uncertainty in the calculated oxygen deficiency ratio is 0.1, assuming an uncertainty of 0.05 at. \% in the measured composition. This measured ratio is lower than what was found for bulk ITO (1.06) by Kim et al.\cite{kimho:intro} and is closer to bulk ITO made by a combination of aquaregia treatment (HNO$_3$, HCl, and distilled water) and dry cleaning with oxygen plasma \cite{kimho:intro}, which was 0.8-0.9. A lower oxygen deficiency in the film may be a result of a decreased number of oxygen vacancies \cite{wu:oxdef}, which may result in a decreased film conductivity \cite{kim:oxdef}.

\begin{figure}
  \begin{center}
    \includegraphics[width=0.4\textwidth]{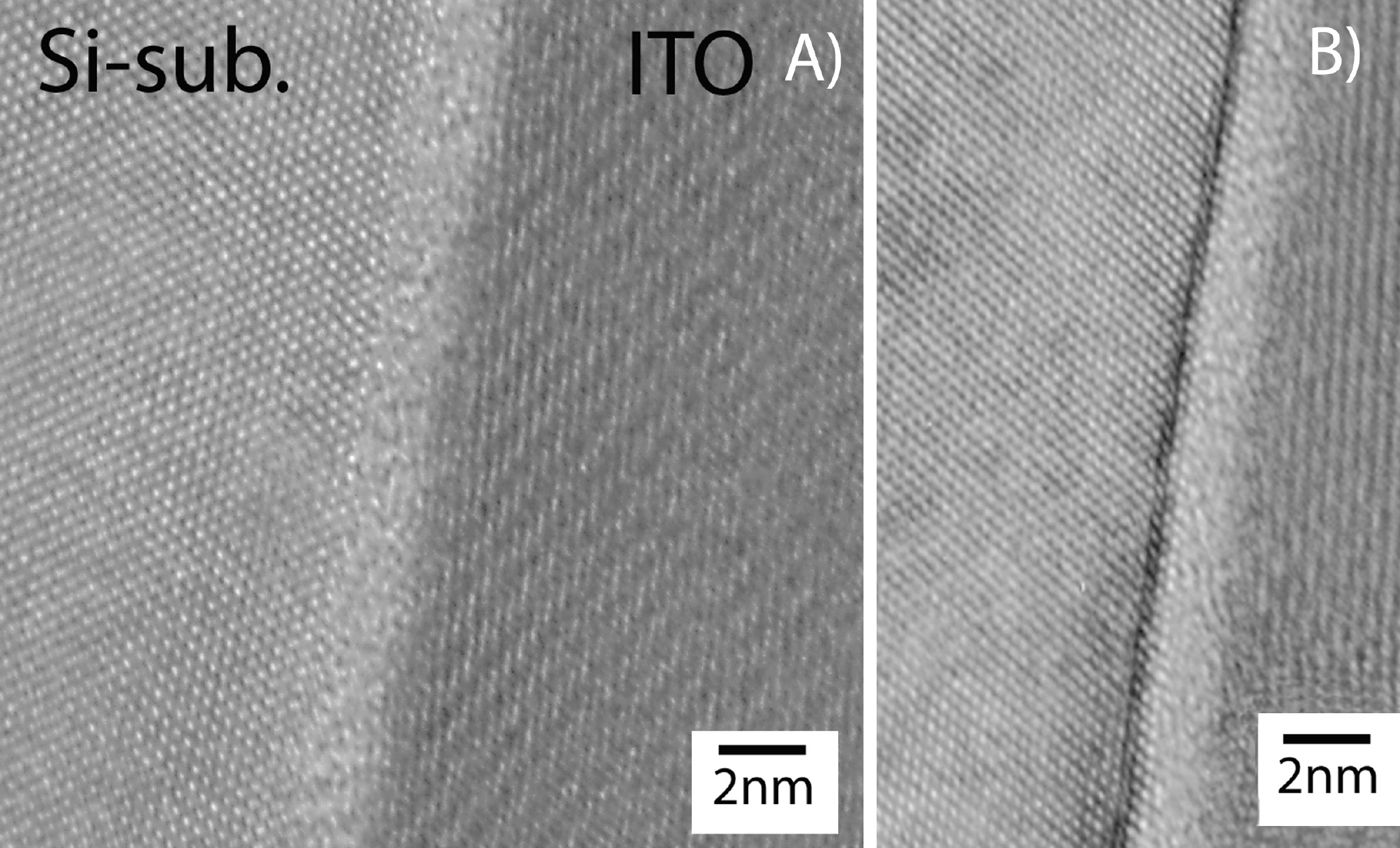}
   \caption{HRTEM images of the interface between the ITO layer and the Si substrate of A) the as deposited and B) the heat treated sample.}
    \label{figure:interface}
  \end{center}
\end{figure}

\begin{figure}
  \begin{center}
    \includegraphics[width=0.4\textwidth]{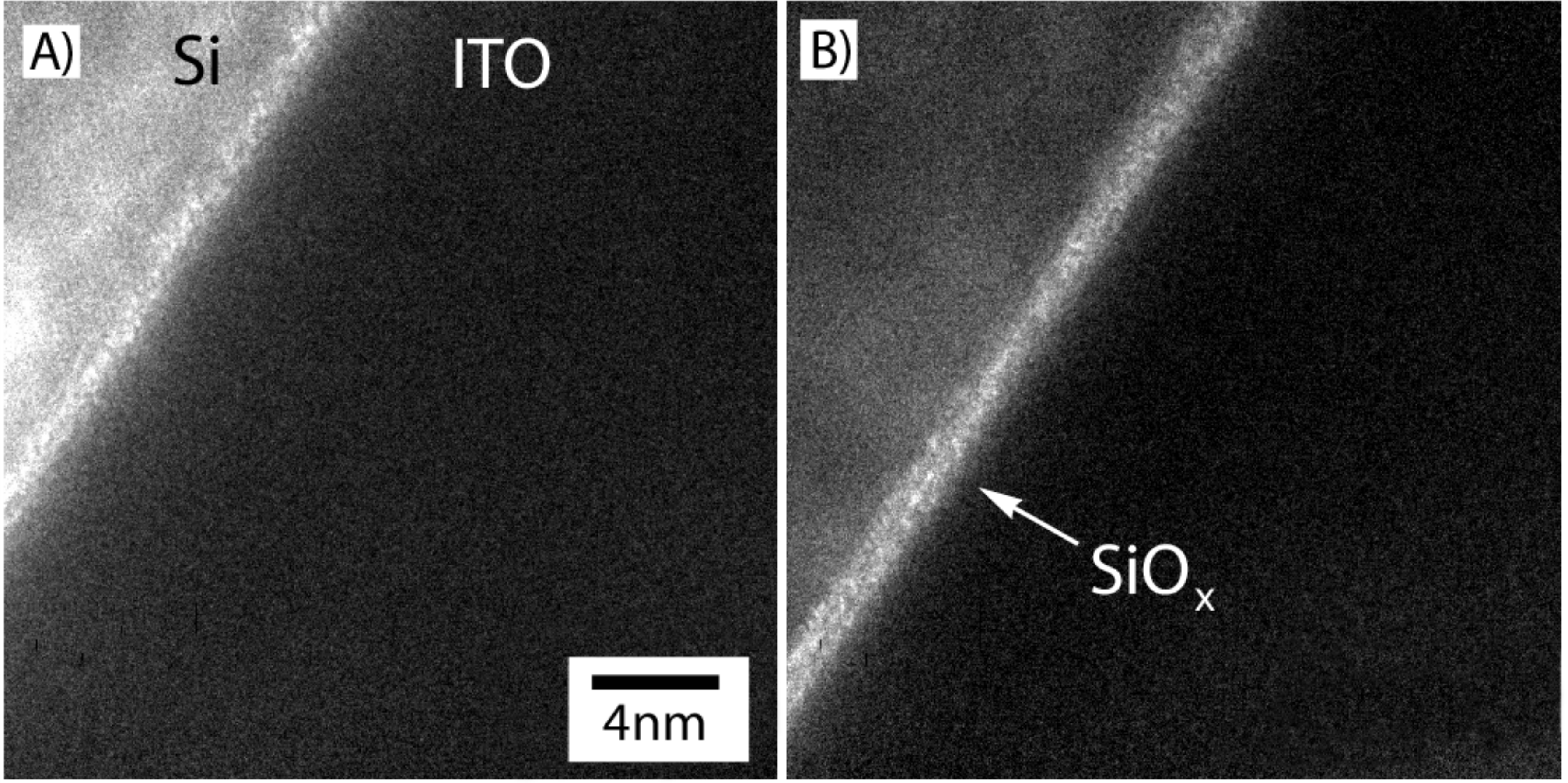}
   \caption{Two EFTEM images from the as deposited sample where A) has been filtered at 16 eV showing the Si and B) at 23 eV showing SiO$_x$.}
    \label{figure:ebeam}
  \end{center}
\end{figure}

\section{Results and Discussion of the ITO/Si interface}
\label{section:itosi}

A detailed HRTEM, EFTEM, and XPS study of the interface is presented in the next sections, together with the effects of electron beam irradiation.  

\begin{table}[hp]
\caption{Binding energy values for ITO film at ITO/Si interface. }
\begin{center}
\noindent
\begin{tabular}{ll}
\hline
\hline
Composition  \, & Binding energy (eV) \, \\
\hline
In$_{I}$ & 444.2 \\
In$_{II}$ & 445.0 \\
In$_{III}$ & 446.4 \\
O$_{I}$ & 530.6 \\
O$_{II}$ & 531.5 \\
O$_{II}$ & 532.7 \\
Sn$_{II}$ & 484.5 \\
Sn$_{II}$ & 486.8 \\
Sn$_{III}$ & 488.2 \\
\hline
\hline
\end{tabular}
\end{center}
\label{table:xpss}
\end{table}

\subsection{Elemental composition measured using XPS}
\label{interfacexx}

Figure \ref{figure:interface} shows HRTEM images of the ITO/Si interface. A 2 nm amorphous layer is visible between the Si substrate and the ITO film. This interface layer is present in both the as deposited and the heat treated sample. Si readily oxidises when exposed to small amounts of O$_2$, and SiO$_x$ is expected to form at the ITO/Si interface. When heat treated at high temperatures, this oxide has been reported to grow \cite{cleva:ITO} (heated at 785 $^\circ$C for 33 min). The interface oxide seen in our samples did not grow during heat treatment. However, our sample has only been heated at 300$^\circ$C for 15 min, and this temperature is probably too low to induce further oxidation.

\begin{table}[hp!]
\caption{Variation of atomic ratios in as deposited sample, with distance from the ITO/Si interface.}
\begin{center}
\noindent
\begin{tabular}{llll}
\hline
\hline
Distance (nm) \, & $\displaystyle \frac{O_I + O_{II}}{In_{total}+Sn_{total}}$ \, & $\displaystyle \frac{O_I}{In_ {II} + Sn_ {II}}$  \, & $\displaystyle \frac{O_{II}}{O_I}$ \,  \\
\hline
11.5 & 1.45 & 1.36 & 0.20 \\
10.0 & 1.45 & 1.28 & 0.25 \\
8.5 & 1.45 & 1.22 & 0.27 \\
7.0 & 1.57 & 1.28 & 0.32 \\
5.5 & 1.62 & 1.11 & 0.63 \\
3.0 & 2.16 & 0.92 & 2.61 \\
1.5 & 3.72 &  &  \\
0.0 & 4.41 &  &  \\
-1.5 & 3.28 &  &  \\
\hline
\hline
\end{tabular}
\end{center}
\label{table:ratioasd}
\end{table}

Ramasse et al.\cite{Quentin:siox} have studied ITO deposited on Si using chemical vapour deposition and re-annealed in vacuum for 30s at 400$^\circ$C. They identified the interfacial oxide as SiO$_x$. EFTEM imaging of the plasmon peak of pure Si (16 eV) and SiO$_x$ (23 eV, which is very close to the plasmon peak of SiO$_2$) was performed to confirm this. Figure \ref{figure:ebeam}A is filtered at 16 eV, which means that the bright areas in the image result from pure Si. Figure \ref{figure:ebeam}B is filtered at 23 eV. The amorphous region is bright when imaging with electrons at 23 eV. As expected, the interface oxide consists of SiO$_x$.

\begin{figure}
  \begin{center}
    \includegraphics[width=0.5\textwidth]{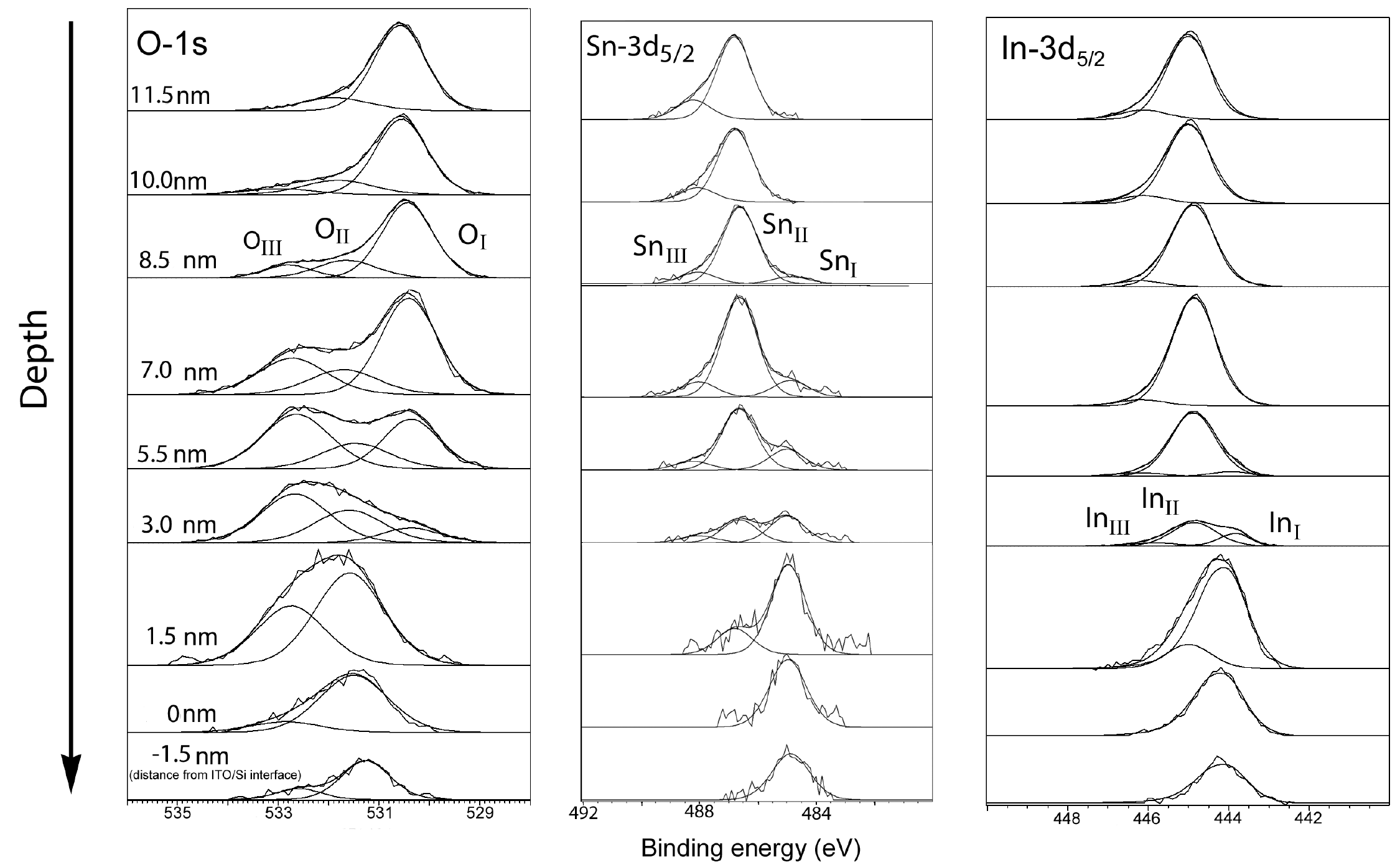}
   \caption{XPS depth profile spectra of the a) O-1s peak, b) Sn-3d$_{5/2}$, and c) In-3d$_{5/2}$ peak of the as deposited sample.}
    \label{figure:xpsasd}
  \end{center}
\end{figure}

\begin{figure}
  \begin{center}
    \includegraphics[width=0.5\textwidth]{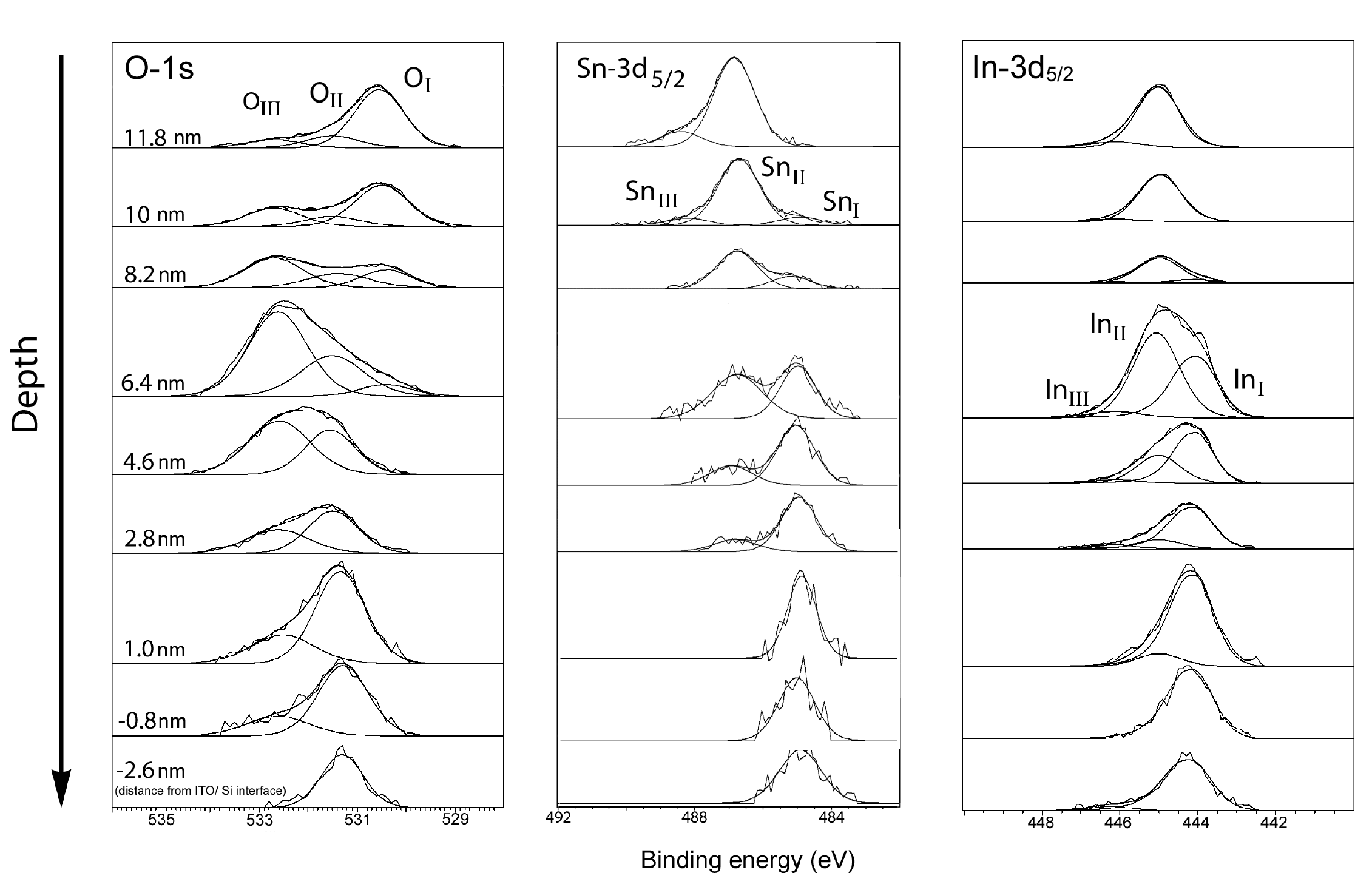}
   \caption{XPS depth profile spectra of the a) O-1s peak, b) Sn-3d$_{5/2}$, and c) In-3d$_{5/2}$ peak of the heat treated sample.}
    \label{figure:xpsht}
  \end{center}
\end{figure}

\begin{figure}
  \begin{center}
    \includegraphics[width=0.5\textwidth]{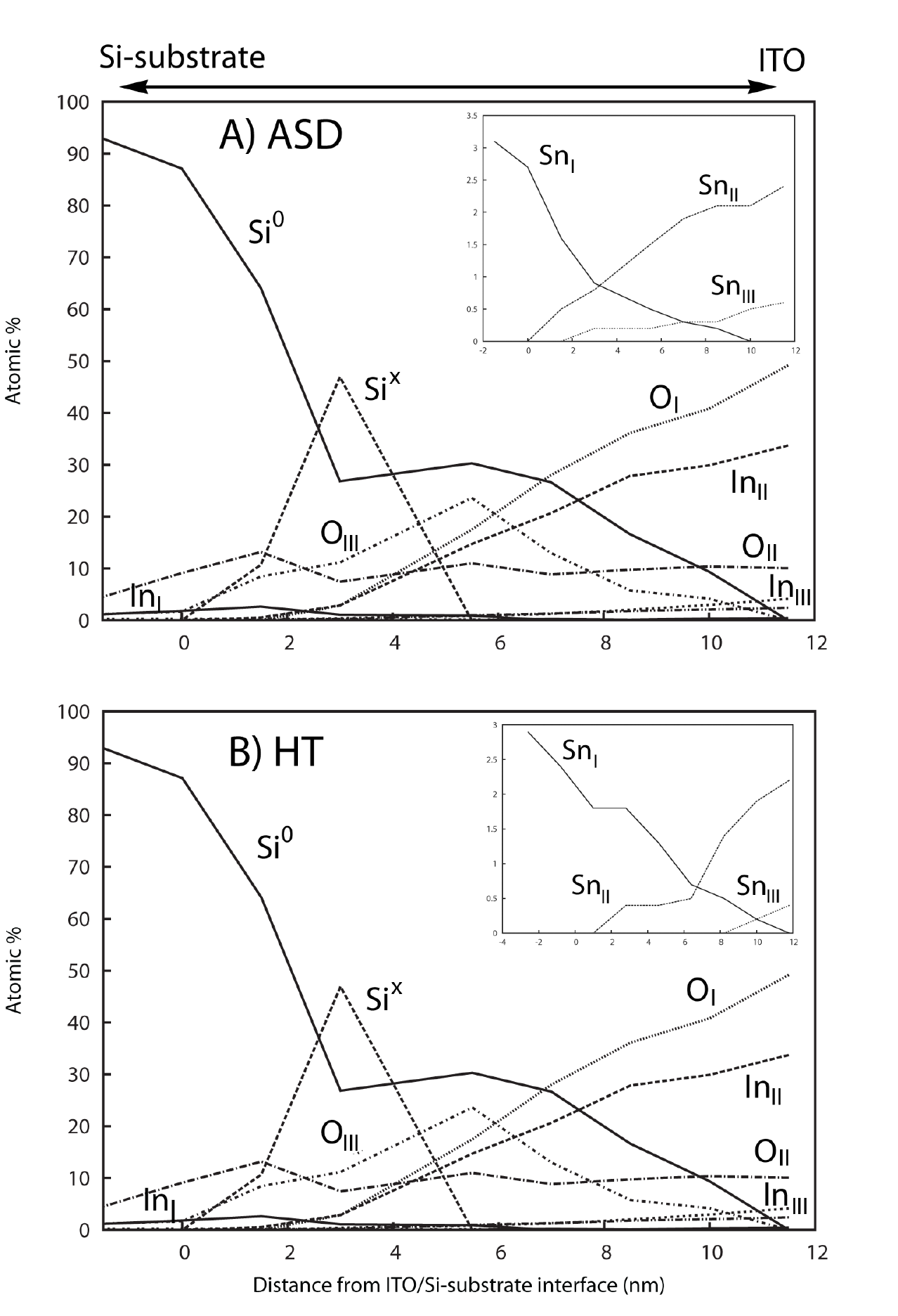}
   \caption{A plot of the atomic percentage of Si, In, Sn, and O near the ITO/Si-Sub interface in the as deposited and heat treated sample.}
    \label{figure:oxPlot}
  \end{center}
\end{figure}

The interface composition of In, O, and Sn was examined using XPS. Figure \ref{figure:xpsasd} and Figure \ref{figure:xpsht} show XPS spectra of the a) O-1s peak, b) Sn-3d peak, and c) In-3d$_{5/2}$ peak near the ITO/Si-substrate interface of the as deposited and heat treated sample, respectively, acquired upon depth profiling using Ar$^+$ etching. The In-3d$_{5/2}$ spectra have been fitted with three components located at 444.2 eV, 445.0 eV, and 446.4 eV. The O-1s spectrum is also fitted with three components at 530.6 eV, 531.5 eV, and 532.7 eV. The Sn-3d$_{5/2}$ peak has been fitted with three components located at a binding energy of 484.9 eV, 486.8 eV, and 488.2 eV. All binding energy values are presented in Table \ref{table:xpss}. The spectra from the as deposited and the heat treated sample are very similar and show that heat treatment did not have a large effect on the elemental and chemical state distribution. From the spectra in Figures \ref{figure:xpsasd} and \ref{figure:xpsht} the atomic percentages of the different oxidation states were measured. These values are plotted in Figure \ref{figure:oxPlot} together with the percentage of Si$^0$ and Si$^x$ (spectra not shown).

Peaks In$_{II}$, In$_{III}$, Sn$_{II}$, Sn$_{III}$, O$_{I}$, and O$_{II}$ are due to crystalline ITO, as discussed in Section \ref{section:bulkxps}. The binding energies of In$_{I}$ and Sn$_{I}$ are similar to the litterature values for pure In and Sn, 443.7 eV\cite{zhua:in} and 485 eV\cite{Ke:sn}, respectively. The pure In and Sn concentration increase towards the interface. If the In$_{III}$, Sn$_{III}$, and O$_{II}$ components exist in crystalline ITO, they should show the same trend (increase or decrease). However, the intensity of the O$_{II}$ component increases towards the ITO/Si-substrate interface where those of In$_{III}$ and Sn$_{III}$ ones decrease. The In$_{III}$ and Sn$_{III}$ peaks could also not be in a composition with the O$_{III}$ peak, for the same reason.

As seen from the HRTEM and EFTEM images in Figures \ref{figure:interface} and \ref{figure:ebeam}, SiO$_x$ is present as a 2 nm layer near the interface. Oxygen in SiO$_2$ has a reported binding energy of 533.05 eV \cite{Hollinger:XPS}. The O$_{III}$ peak has a binding energy of 532.7 eV and appears around the same depth as the Si-2p peak (spectra not shown). This suggests that the O$_{III}$ peak results from oxygen in SiO$_x$. The peak is present at a distance of 8 nm from the interface. This distance is more than what was identified as SiO$_x$ by TEM and EFTEM. This is because the image resolution in the TEM (0.19 nm) is far better than the depth resolution in XPS (which is the photoelectron escape depth of about 7 nm).

The electroneutrality principle was again used to determine the chemical state of the oxide close to the interface, for both the as deposited and heat treated sample. Since both TEM and XPS data show no significant differences between the two samples, only the data of the as deposited sample is presented in Table \ref{table:ratioasd}. The (O$_{I}$+O$_{II}$)/(In$_{total}$+Sn$_{total}$) ratio is within 1.4-1.6 for the first four spectra in Figure \ref{figure:xpsasd}. This means that the oxide probably contains mostly O$^{2-}$. Towards the interface the ratio increases to around 4. This is considerably higher than what was observed for non-interfacial ITO in Section \ref{section:bulkxps}.

As discussed in Section \ref{section:bulkxps}, the O$_{II}$/O$_{I}$ ratio may be used to determine the oxygen deficiency in the material \cite{kimho:intro}. The oxygen ratio is presented in Table \ref{table:ratioasd} for the as deposited sample. The oxygen deficiency ratio of non-interfacial ITO as presented in Section \ref{section:bulkxps} was 0.2 and 0.21 for the as deposited and the heat treated sample, respectively. The ratio of oxygen deficient regions near the interface is higher. At 11 nm from the interface the ratio is the same as bulk for ITO film. Moving towards the interface the ratio increases, and at the interface the ratio is close to 2. These results show that there is presence of pure In$^0$ and Sn$^0$ at the interface, together with highly oxygen deficient regions. 

The pure In and Sn at the interface have most likely formed during deposition, when oxygen from ITO may have reacted with the unsaturated Si at the wafer surface to make SiO$_x$ \cite{diplas:ITO}. Pure In and Sn have therefore probably been a bi-product of this reaction.

\subsection{Electron beam induced crystallization}

\begin{figure}
  \begin{center}
    \includegraphics[width=0.5\textwidth]{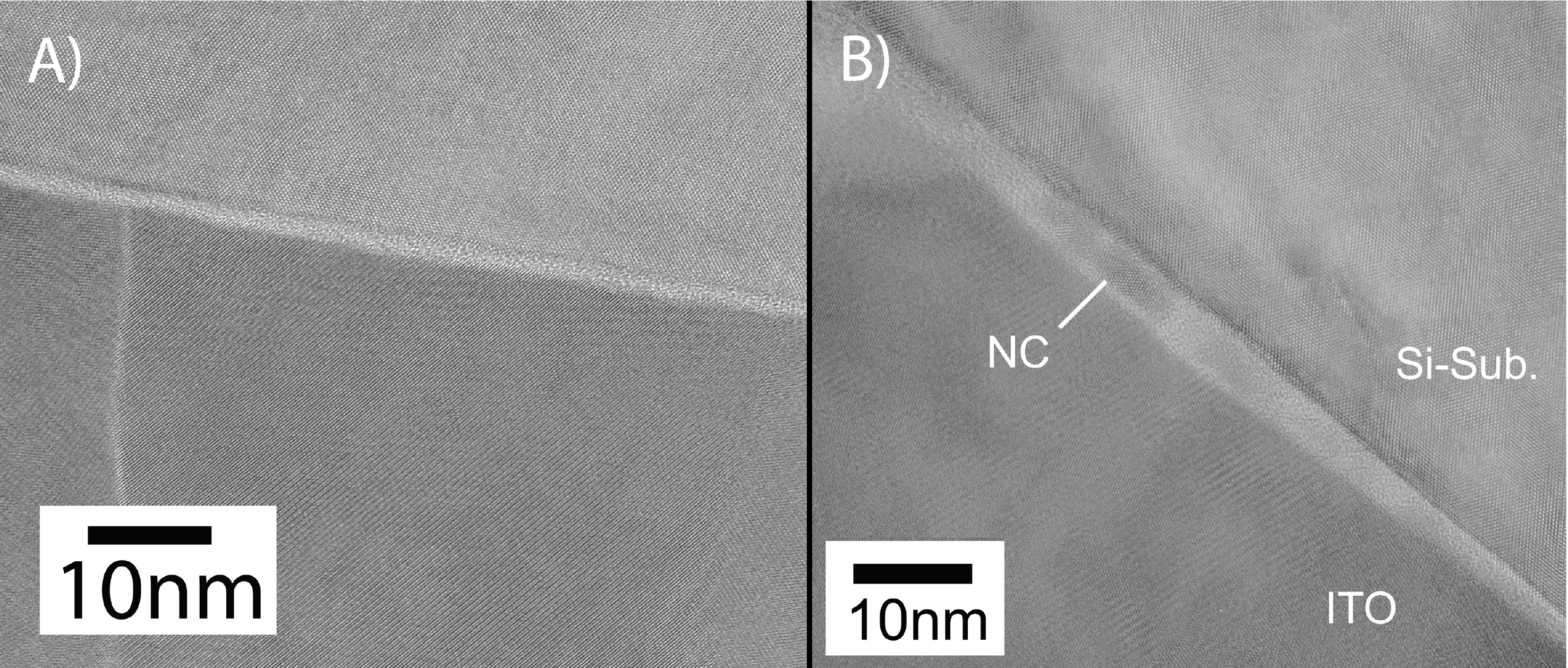}
   \caption{HRTEM images of the ITO/Si interface, (A) before  and (B) after electron beam exposure.}
    \label{figure:eee}
  \end{center}
\end{figure}

In the previous section XPS analysis indicated that pure In and Sn co-exist with oxygen deficient regions near the interface. Since conventional TEM bright field imaging did not reveal presence of In clusters, one can assume that pure In exists as amorphous nanoclusters near the ITO/Si interface. Upon exposure to high electron beam illumination, the sample at the interface reacted with the electron beam. Figure \ref{figure:eee} shows an HRTEM image of the as deposited sample before (A) and after (B) electron beam exposure. During electron beam illumination, small nanocrystals appear at the ITO/Si interface. 

\begin{figure*}
  \begin{center}
    \includegraphics[width=0.7\textwidth]{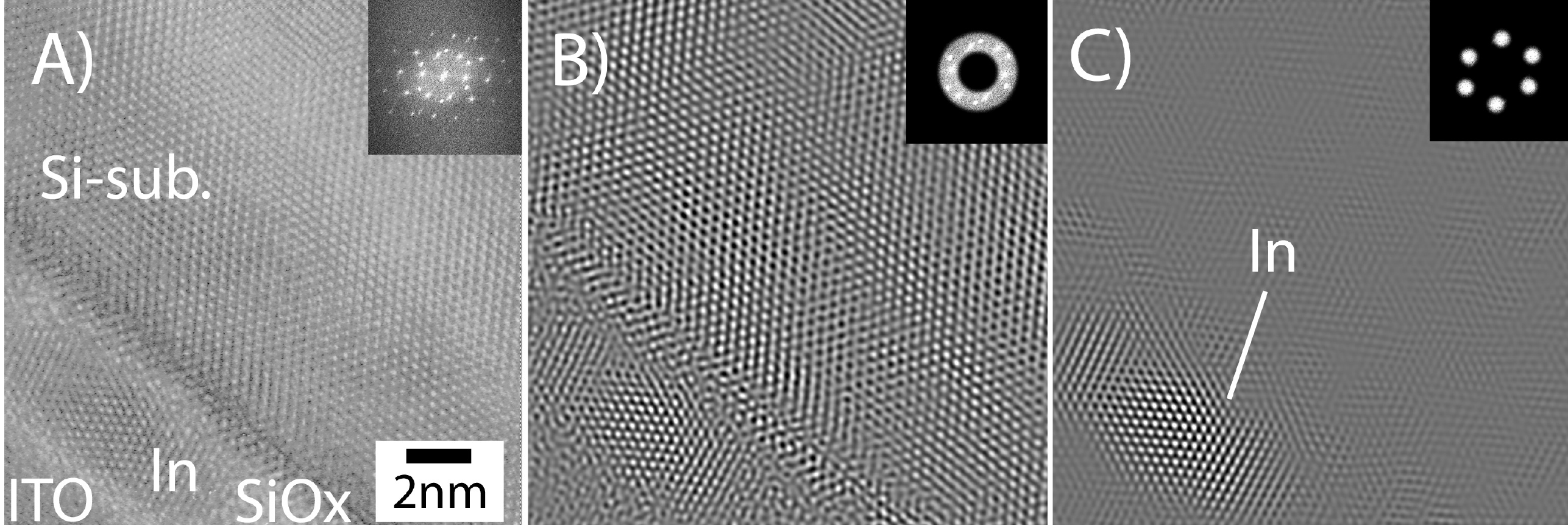}
   \caption{HRTEM images of the Si substrate and the interface nanocrystal, where A) is the HRTEM image, B) is the HRTEM image after applying a circular mask, and C) is after applying a mask of only the In diffraction pattern.}
    \label{figure:mask}
  \end{center}
\end{figure*}

Figure \ref{figure:mask}A shows HRTEM images of the ITO/Si interface after only seconds of electron beam irradiation. After a short interval of electron irradiation, nanocrystals appear at the ITO/Si substrate interface in the interfacial oxide layer. The Fast Fourier Transformed (FFT) image inserted in the Figure, shows the diffraction pattern of both the Si substrate and the interface nanocrystal. Figure \ref{figure:mask}B shows the HRTEM image after applying a circular mask with both the Si and the interface nanocrystal diffraction pattern, while Figure \ref{figure:mask}C has resulted from applying a mask of only the diffraction patterns of the interface nanocrystal. The inserted FFT patterns in the image show the applied mask. After calibrating the diffraction pattern using the Si diffraction pattern as a reference, the lattice parameters of the interface nanocrystals were found to be 0.270 nm and 0.243 nm. In has a space group of I4/mmm (No. 139), with $a=b= 0.325$ nm and $c= 0.495$ nm \cite{In:ref}. The measured lattice parameters fit well with the (101) and (002) plane of pure In, which are 0.270 nm and 0.244 nm, respectively. Therefore, the structure of the interface nanocrystal fits well with the structure of pure In. The elemental composition near the interface as discussed in the beginning of this section was found to be composed of SiO$_x$ in addition to the presence of In and Sn nanoclusters. The TEM results strengthen the argument that pure In is present as nanoclusters. During high energy electron irradiation, the In clusters crystallize. In addition to this crystallization, the interface oxide increases in thickness from 2 nm to 5 nm (see Figure \ref{figure:mask}).

Pure In at the interface has previously been reported by Ow-Yang et al. \cite{cleva:ITO} and Kobayashi et al.\cite{intro:NC}. Kobayashi et al.\cite{intro:NC} found presence of metallic In by XRD when the deposition of ITO occurred at $0^\circ$ angle. Ow-Yang et al. \cite{cleva:ITO} studied the ITO/Si interface by XRD and thermodynamics. Even though no other reaction product was detected by TEM at the ITO/Si interface, pure In was observed by XPS. This reaction was also found to be thermodynamically stable at 1058K, as shown below\cite{cleva:ITO}.   

\[2\text{In}_2\text{O}_{3(s)} + 3\text{Si}_{(s)} = 4\text{In}_{(s)} + 3\text{SiO}_{2(s)}
\]

\[\Delta \text{G} = -981.3 \text{kJ/mol}
\]

The In crystallization could be attributed to localized heating at the interface or to charging produced by the electron beam. An in-situ heating experiment of the sample showed no nanocrystal formation at the ITO/Si interface. It is therefore likely that crystallization and growth of the In nanocrystals are due to direct electron beam induced splitting and re-configuration of atomic bonds. Since In is present as In$^{3+}$, the electron beam might attract additional In to the interface, thereby resulting in larger In clusters. This may be assisted by the oxygen depletion at the interfacial areas and subsequently by the presence of broken In-O bonds. The crystallization may naturally occur when the In cluster reaches its critical radius. Nanocluster formation and e-beam induced crystallization may have significant implications in manipulating the ITO/Si interface at the nano-scale via a combined method of deposition and electron beam irradiation in the early stages of deposition.

\section{Conclusion}

As deposited and heat treated ITO films made by sputter deposition were studied in detail using TEM and XPS. The samples were very similar and both crystalline. This was attributed to plasma enhanced crystallization during deposition. The non-interfacial ITO consists only of crystalline ITO. XPS depth profiling and TEM of the ITO/Si interface revealed increasing amounts of In and Sn towards the interface, as well as the presence of an SiO$_x$ layer. This interface oxide did not grow during heat treatment. Pure In was present as amorphous clusters. Close to the interface areas oxygen deficient regions were also found. During electron beam exposure of the interface, the In nanoclusters crystallize and grow. In addition, growth of the SiO$_x$ layer occurred. 


\begin{thebibliography}{33}
\expandafter\ifx\csname natexlab\endcsname\relax\def\natexlab#1{#1}\fi
\expandafter\ifx\csname bibnamefont\endcsname\relax
  \def\bibnamefont#1{#1}\fi
\expandafter\ifx\csname bibfnamefont\endcsname\relax
  \def\bibfnamefont#1{#1}\fi
\expandafter\ifx\csname citenamefont\endcsname\relax
  \def\citenamefont#1{#1}\fi
\expandafter\ifx\csname url\endcsname\relax
  \def\url#1{\texttt{#1}}\fi
\expandafter\ifx\csname urlprefix\endcsname\relax\def\urlprefix{URL }\fi
\providecommand{\bibinfo}[2]{#2}
\providecommand{\eprint}[2][]{\url{#2}}

\bibitem[{\citenamefont{Xirouchaki et~al.}(1996)\citenamefont{Xirouchaki,
  Kiriakidis, Pedersen, and Fritzsche}}]{xirouchaki:intro}
\bibinfo{author}{\bibfnamefont{C.}~\bibnamefont{Xirouchaki}},
  \bibinfo{author}{\bibfnamefont{G.}~\bibnamefont{Kiriakidis}},
  \bibinfo{author}{\bibfnamefont{T.}~\bibnamefont{Pedersen}}, \bibnamefont{and}
  \bibinfo{author}{\bibfnamefont{H.}~\bibnamefont{Fritzsche}},
  \bibinfo{journal}{J. of Appl. Phys.} \textbf{\bibinfo{volume}{79}},
  \bibinfo{pages}{9349} (\bibinfo{year}{1996}).

\bibitem[{\citenamefont{Kim et~al.}(2005)\citenamefont{Kim, Park, Kim, and
  Sung}}]{kin:intro}
\bibinfo{author}{\bibfnamefont{S.~H.} \bibnamefont{Kim}},
  \bibinfo{author}{\bibfnamefont{N.-M.} \bibnamefont{Park}},
  \bibinfo{author}{\bibfnamefont{T.}~\bibnamefont{Kim}}, \bibnamefont{and}
  \bibinfo{author}{\bibfnamefont{G.}~\bibnamefont{Sung}},
  \bibinfo{journal}{Thin Solid Films} \textbf{\bibinfo{volume}{475}},
  \bibinfo{pages}{262} (\bibinfo{year}{2005}).

\bibitem[{\citenamefont{Ando et~al.}(2003)\citenamefont{Ando, Nishimura,
  Onisawa, and Minemura}}]{ando:intro}
\bibinfo{author}{\bibfnamefont{M.}~\bibnamefont{Ando}},
  \bibinfo{author}{\bibfnamefont{E.}~\bibnamefont{Nishimura}},
  \bibinfo{author}{\bibfnamefont{K.}~\bibnamefont{Onisawa}}, \bibnamefont{and}
  \bibinfo{author}{\bibfnamefont{T.}~\bibnamefont{Minemura}},
  \bibinfo{journal}{J. of Appl. Phys.} \textbf{\bibinfo{volume}{93}},
  \bibinfo{pages}{1032} (\bibinfo{year}{2003}).

\bibitem[{\citenamefont{Chopra et~al.}(2000)\citenamefont{Chopra, Major, and
  Pandya}}]{chopra:intro}
\bibinfo{author}{\bibfnamefont{K.}~\bibnamefont{Chopra}},
  \bibinfo{author}{\bibfnamefont{S.}~\bibnamefont{Major}}, \bibnamefont{and}
  \bibinfo{author}{\bibfnamefont{K.}~\bibnamefont{Pandya}},
  \bibinfo{journal}{Thin Solid Films} \textbf{\bibinfo{volume}{102}},
  \bibinfo{pages}{1} (\bibinfo{year}{2000}).

\bibitem[{\citenamefont{Ohta et~al.}(2000)\citenamefont{Ohta, Orita, Hirano,
  Tanji, Kawazoe, and Hosono}}]{otha:intro}
\bibinfo{author}{\bibfnamefont{H.}~\bibnamefont{Ohta}},
  \bibinfo{author}{\bibfnamefont{M.}~\bibnamefont{Orita}},
  \bibinfo{author}{\bibfnamefont{M.}~\bibnamefont{Hirano}},
  \bibinfo{author}{\bibfnamefont{H.}~\bibnamefont{Tanji}},
  \bibinfo{author}{\bibfnamefont{H.}~\bibnamefont{Kawazoe}}, \bibnamefont{and}
  \bibinfo{author}{\bibfnamefont{H.}~\bibnamefont{Hosono}},
  \bibinfo{journal}{Appl. Phys. Lett.} \textbf{\bibinfo{volume}{76}},
  \bibinfo{pages}{2740} (\bibinfo{year}{2000}).

\bibitem[{\citenamefont{Tahar et~al.}(1998)\citenamefont{Tahar, Ban, Ohaya, and
  Takahashi}}]{tahar:intro}
\bibinfo{author}{\bibfnamefont{R.}~\bibnamefont{Tahar}},
  \bibinfo{author}{\bibfnamefont{T.}~\bibnamefont{Ban}},
  \bibinfo{author}{\bibfnamefont{Y.}~\bibnamefont{Ohaya}}, \bibnamefont{and}
  \bibinfo{author}{\bibfnamefont{Y.}~\bibnamefont{Takahashi}},
  \bibinfo{journal}{J. Appl. Phys.} \textbf{\bibinfo{volume}{83}},
  \bibinfo{pages}{2631} (\bibinfo{year}{1998}).

\bibitem[{\citenamefont{Kobayashi et~al.}(1991)\citenamefont{Kobayashi, Ishida,
  Nakato, and Tsubomura}}]{kobayashi:intro}
\bibinfo{author}{\bibfnamefont{H.}~\bibnamefont{Kobayashi}},
  \bibinfo{author}{\bibfnamefont{T.}~\bibnamefont{Ishida}},
  \bibinfo{author}{\bibfnamefont{Y.}~\bibnamefont{Nakato}}, \bibnamefont{and}
  \bibinfo{author}{\bibfnamefont{H.}~\bibnamefont{Tsubomura}},
  \bibinfo{journal}{J. Appl. Phys.} \textbf{\bibinfo{volume}{69}},
  \bibinfo{pages}{1736} (\bibinfo{year}{1991}).

\bibitem[{\citenamefont{Kim et~al.}(2003)\citenamefont{Kim, Park, Ansari, Lee,
  and Shin}}]{Soonkim:in}
\bibinfo{author}{\bibfnamefont{Y.-S.} \bibnamefont{Kim}},
  \bibinfo{author}{\bibfnamefont{Y.-C.} \bibnamefont{Park}},
  \bibinfo{author}{\bibfnamefont{S.}~\bibnamefont{Ansari}},
  \bibinfo{author}{\bibfnamefont{B.-S.} \bibnamefont{Lee}}, \bibnamefont{and}
  \bibinfo{author}{\bibfnamefont{H.-S.} \bibnamefont{Shin}},
  \bibinfo{journal}{Thin Solid Films} \textbf{\bibinfo{volume}{426}},
  \bibinfo{pages}{124} (\bibinfo{year}{2003}).

\bibitem[{\citenamefont{Kim et~al.}(1999{\natexlab{a}})\citenamefont{Kim, Ho,
  Thomas, Friend, Cacialli, and Bao}}]{kimho:intro}
\bibinfo{author}{\bibfnamefont{J.}~\bibnamefont{Kim}},
  \bibinfo{author}{\bibfnamefont{P.}~\bibnamefont{Ho}},
  \bibinfo{author}{\bibfnamefont{D.}~\bibnamefont{Thomas}},
  \bibinfo{author}{\bibfnamefont{R.}~\bibnamefont{Friend}},
  \bibinfo{author}{\bibfnamefont{F.}~\bibnamefont{Cacialli}}, \bibnamefont{and}
  \bibinfo{author}{\bibfnamefont{G.}~\bibnamefont{Bao}},
  \bibinfo{journal}{Chem. Phys. Lett.} \textbf{\bibinfo{volume}{315}},
  \bibinfo{pages}{307} (\bibinfo{year}{1999}{\natexlab{a}}).

\bibitem[{\citenamefont{Ishida et~al.}(1993)\citenamefont{Ishida, Kobayashi,
  and Nakato}}]{ishida:o}
\bibinfo{author}{\bibfnamefont{T.}~\bibnamefont{Ishida}},
  \bibinfo{author}{\bibfnamefont{H.}~\bibnamefont{Kobayashi}},
  \bibnamefont{and} \bibinfo{author}{\bibfnamefont{Y.}~\bibnamefont{Nakato}},
  \bibinfo{journal}{J. Appl. Phys.} \textbf{\bibinfo{volume}{73}},
  \bibinfo{pages}{4344} (\bibinfo{year}{1993}).

\bibitem[{\citenamefont{Fan and Goodenough}(1977)}]{fan:ox}
\bibinfo{author}{\bibfnamefont{J.}~\bibnamefont{Fan}} \bibnamefont{and}
  \bibinfo{author}{\bibfnamefont{J.}~\bibnamefont{Goodenough}},
  \bibinfo{journal}{J. Appl. Phys.} \textbf{\bibinfo{volume}{48}},
  \bibinfo{pages}{3524} (\bibinfo{year}{1977}).

\bibitem[{\citenamefont{Ow-Yang et~al.}(2000)\citenamefont{Ow-Yang, Shigesato,
  and Paine}}]{cleva:ITO}
\bibinfo{author}{\bibfnamefont{C.~W.} \bibnamefont{Ow-Yang}},
  \bibinfo{author}{\bibfnamefont{Y.}~\bibnamefont{Shigesato}},
  \bibnamefont{and} \bibinfo{author}{\bibfnamefont{D.~C.} \bibnamefont{Paine}},
  \bibinfo{journal}{J. of Appl. Phys.} \textbf{\bibinfo{volume}{88}},
  \bibinfo{pages}{3717} (\bibinfo{year}{2000}).

\bibitem[{\citenamefont{Ramasse et~al.}(2009)\citenamefont{Ramasse, Anapolsky,
  Lazik, and Wang}}]{Quentin:siox}
\bibinfo{author}{\bibfnamefont{Q.}~\bibnamefont{Ramasse}},
  \bibinfo{author}{\bibfnamefont{A.}~\bibnamefont{Anapolsky}},
  \bibinfo{author}{\bibfnamefont{C.}~\bibnamefont{Lazik}}, \bibnamefont{and}
  \bibinfo{author}{\bibfnamefont{M.~J. A. A.~A.} \bibnamefont{Wang}},
  \bibinfo{journal}{J. of Appl. Phys.} \textbf{\bibinfo{volume}{105}},
  \bibinfo{pages}{033716} (\bibinfo{year}{2009}).

\bibitem[{\citenamefont{Kobayashi et~al.}(1992)\citenamefont{Kobayashi, Ishida,
  Nakamura, Nakato, and Tsubomura}}]{intro:NC}
\bibinfo{author}{\bibfnamefont{H.}~\bibnamefont{Kobayashi}},
  \bibinfo{author}{\bibfnamefont{T.}~\bibnamefont{Ishida}},
  \bibinfo{author}{\bibfnamefont{K.}~\bibnamefont{Nakamura}},
  \bibinfo{author}{\bibfnamefont{Y.}~\bibnamefont{Nakato}}, \bibnamefont{and}
  \bibinfo{author}{\bibfnamefont{H.}~\bibnamefont{Tsubomura}},
  \bibinfo{journal}{J. Appl. Phys.} \textbf{\bibinfo{volume}{72}},
  \bibinfo{pages}{5288} (\bibinfo{year}{1992}).

\bibitem[{\citenamefont{Zhua et~al.}(2000)\citenamefont{Zhua, Huan, Zhang, and
  Wee}}]{zhua:in}
\bibinfo{author}{\bibfnamefont{F.}~\bibnamefont{Zhua}},
  \bibinfo{author}{\bibfnamefont{C.}~\bibnamefont{Huan}},
  \bibinfo{author}{\bibfnamefont{K.}~\bibnamefont{Zhang}}, \bibnamefont{and}
  \bibinfo{author}{\bibfnamefont{A.}~\bibnamefont{Wee}}, \bibinfo{journal}{Thin
  Solid Films} \textbf{\bibinfo{volume}{359}}, \bibinfo{pages}{244}
  (\bibinfo{year}{2000}).

\bibitem[{\citenamefont{de~Carvalho et~al.}(2000)\citenamefont{de~Carvalho,
  do~Rego, Amaral, Brogueira, and Lavareda}}]{xps:in}
\bibinfo{author}{\bibfnamefont{C.~N.} \bibnamefont{de~Carvalho}},
  \bibinfo{author}{\bibfnamefont{A.~B.} \bibnamefont{do~Rego}},
  \bibinfo{author}{\bibfnamefont{A.}~\bibnamefont{Amaral}},
  \bibinfo{author}{\bibfnamefont{P.}~\bibnamefont{Brogueira}},
  \bibnamefont{and} \bibinfo{author}{\bibfnamefont{G.}~\bibnamefont{Lavareda}},
  \bibinfo{journal}{Surface and Coatings Technology}
  \textbf{\bibinfo{volume}{124}}, \bibinfo{pages}{70} (\bibinfo{year}{2000}).

\bibitem[{\citenamefont{Jimenez et~al.}(1996)\citenamefont{Jimenez, Fernandez,
  Espinos, and Gonzalez-Elipe}}]{Jimenez:sn}
\bibinfo{author}{\bibfnamefont{V.}~\bibnamefont{Jimenez}},
  \bibinfo{author}{\bibfnamefont{A.}~\bibnamefont{Fernandez}},
  \bibinfo{author}{\bibfnamefont{J.}~\bibnamefont{Espinos}}, \bibnamefont{and}
  \bibinfo{author}{\bibfnamefont{A.}~\bibnamefont{Gonzalez-Elipe}},
  \bibinfo{journal}{Surf. Sci.} \textbf{\bibinfo{volume}{350}},
  \bibinfo{pages}{123} (\bibinfo{year}{1996}).

\bibitem[{\citenamefont{Ke et~al.}(2007)\citenamefont{Ke, Huang, Cai, and
  Sun}}]{Ke:sn}
\bibinfo{author}{\bibfnamefont{F.-S.} \bibnamefont{Ke}},
  \bibinfo{author}{\bibfnamefont{L.}~\bibnamefont{Huang}},
  \bibinfo{author}{\bibfnamefont{J.-S.} \bibnamefont{Cai}}, \bibnamefont{and}
  \bibinfo{author}{\bibfnamefont{S.-G.} \bibnamefont{Sun}},
  \bibinfo{journal}{Electrochimica Acta 52} \textbf{\bibinfo{volume}{52}},
  \bibinfo{pages}{6741} (\bibinfo{year}{2007}).

\bibitem[{\citenamefont{Peng et~al.}(2002)\citenamefont{Peng, Meng, Wang, Wang,
  Zhang, Liu, and Zhang}}]{oxdef:1}
\bibinfo{author}{\bibfnamefont{X.~S.} \bibnamefont{Peng}},
  \bibinfo{author}{\bibfnamefont{G.~W.} \bibnamefont{Meng}},
  \bibinfo{author}{\bibfnamefont{X.~F.} \bibnamefont{Wang}},
  \bibinfo{author}{\bibfnamefont{Y.~W.} \bibnamefont{Wang}},
  \bibinfo{author}{\bibfnamefont{J.}~\bibnamefont{Zhang}},
  \bibinfo{author}{\bibfnamefont{X.}~\bibnamefont{Liu}}, \bibnamefont{and}
  \bibinfo{author}{\bibfnamefont{L.~D.} \bibnamefont{Zhang}},
  \bibinfo{journal}{Chem. Mater.} \textbf{\bibinfo{volume}{14}},
  \bibinfo{pages}{4490} (\bibinfo{year}{2002}).

\bibitem[{\citenamefont{http://www.casaxps.com/}()}]{casa:xps}
\bibinfo{author}{\bibnamefont{http://www.casaxps.com/}}.

\bibitem[{\citenamefont{Diplas et~al.}(2007)\citenamefont{Diplas, Ulyashin,
  Maknys, Gunnaes, J{\o }rgensen, Wright, Watts, Olsen, and
  Finstad}}]{spyros:2}
\bibinfo{author}{\bibfnamefont{S.}~\bibnamefont{Diplas}},
  \bibinfo{author}{\bibfnamefont{A.}~\bibnamefont{Ulyashin}},
  \bibinfo{author}{\bibfnamefont{K.}~\bibnamefont{Maknys}},
  \bibinfo{author}{\bibfnamefont{A.}~\bibnamefont{Gunnaes}},
  \bibinfo{author}{\bibfnamefont{S.}~\bibnamefont{J{\o }rgensen}},
  \bibinfo{author}{\bibfnamefont{D.}~\bibnamefont{Wright}},
  \bibinfo{author}{\bibfnamefont{J.}~\bibnamefont{Watts}},
  \bibinfo{author}{\bibfnamefont{A.}~\bibnamefont{Olsen}}, \bibnamefont{and}
  \bibinfo{author}{\bibfnamefont{T.}~\bibnamefont{Finstad}},
  \bibinfo{journal}{Thin Solid Films} \textbf{\bibinfo{volume}{515}},
  \bibinfo{pages}{8539} (\bibinfo{year}{2007}).

\bibitem[{\citenamefont{Song et~al.}(1998)\citenamefont{Song, Shigesato, Yasui,
  Ow-Yang, and Paine}}]{M:1}
\bibinfo{author}{\bibfnamefont{P.~K.} \bibnamefont{Song}},
  \bibinfo{author}{\bibfnamefont{Y.}~\bibnamefont{Shigesato}},
  \bibinfo{author}{\bibfnamefont{I.}~\bibnamefont{Yasui}},
  \bibinfo{author}{\bibfnamefont{C.~W.} \bibnamefont{Ow-Yang}},
  \bibnamefont{and} \bibinfo{author}{\bibfnamefont{D.~C.} \bibnamefont{Paine}},
  \bibinfo{journal}{Japanese J. of Appl. Phys.} \textbf{\bibinfo{volume}{37}},
  \bibinfo{pages}{1870} (\bibinfo{year}{1998}).

\bibitem[{\citenamefont{Song et~al.}(1999{\natexlab{a}})\citenamefont{Song,
  Shigesato, Kamei, and Yasui}}]{M:2}
\bibinfo{author}{\bibfnamefont{P.~K.} \bibnamefont{Song}},
  \bibinfo{author}{\bibfnamefont{Y.}~\bibnamefont{Shigesato}},
  \bibinfo{author}{\bibfnamefont{M.}~\bibnamefont{Kamei}}, \bibnamefont{and}
  \bibinfo{author}{\bibfnamefont{I.}~\bibnamefont{Yasui}},
  \bibinfo{journal}{Japanese J. of Appl. Phys.} \textbf{\bibinfo{volume}{38}},
  \bibinfo{pages}{2921} (\bibinfo{year}{1999}{\natexlab{a}}).

\bibitem[{\citenamefont{Antony et~al.}(2004)\citenamefont{Antony, Nisha, Manoj,
  and Jayaraj}}]{M:3}
\bibinfo{author}{\bibfnamefont{A.}~\bibnamefont{Antony}},
  \bibinfo{author}{\bibfnamefont{M.}~\bibnamefont{Nisha}},
  \bibinfo{author}{\bibfnamefont{R.}~\bibnamefont{Manoj}}, \bibnamefont{and}
  \bibinfo{author}{\bibfnamefont{M.~K.} \bibnamefont{Jayaraj}},
  \bibinfo{journal}{Appl. Surf. Sci.} \textbf{\bibinfo{volume}{225}},
  \bibinfo{pages}{294} (\bibinfo{year}{2004}).

\bibitem[{\citenamefont{Mei-Zhen et~al.}(2008)\citenamefont{Mei-Zhen, Job,
  De-Sheng, and Fahrner}}]{M:4}
\bibinfo{author}{\bibfnamefont{G.}~\bibnamefont{Mei-Zhen}},
  \bibinfo{author}{\bibfnamefont{R.}~\bibnamefont{Job}},
  \bibinfo{author}{\bibfnamefont{X.}~\bibnamefont{De-Sheng}}, \bibnamefont{and}
  \bibinfo{author}{\bibfnamefont{W.~R.} \bibnamefont{Fahrner}},
  \bibinfo{journal}{Chinese Phys. Lett.} \textbf{\bibinfo{volume}{25}},
  \bibinfo{pages}{1380} (\bibinfo{year}{2008}).

\bibitem[{\citenamefont{Song et~al.}(1999{\natexlab{b}})\citenamefont{Song,
  Akao, Kamei, Shigesato, and Yasui}}]{ito:temp1}
\bibinfo{author}{\bibfnamefont{P.~K.} \bibnamefont{Song}},
  \bibinfo{author}{\bibfnamefont{H.}~\bibnamefont{Akao}},
  \bibinfo{author}{\bibfnamefont{M.}~\bibnamefont{Kamei}},
  \bibinfo{author}{\bibfnamefont{Y.}~\bibnamefont{Shigesato}},
  \bibnamefont{and} \bibinfo{author}{\bibfnamefont{I.}~\bibnamefont{Yasui}},
  \bibinfo{journal}{Japanese J. of Appl. Phys.} \textbf{\bibinfo{volume}{38}},
  \bibinfo{pages}{5224} (\bibinfo{year}{1999}{\natexlab{b}}).

\bibitem[{\citenamefont{Paine et~al.}(1999)\citenamefont{Paine, Whitson,
  Janiac, Beresford, and Yang}}]{ito:temp2}
\bibinfo{author}{\bibfnamefont{D.~C.} \bibnamefont{Paine}},
  \bibinfo{author}{\bibfnamefont{T.}~\bibnamefont{Whitson}},
  \bibinfo{author}{\bibfnamefont{D.}~\bibnamefont{Janiac}},
  \bibinfo{author}{\bibfnamefont{R.}~\bibnamefont{Beresford}},
  \bibnamefont{and} \bibinfo{author}{\bibfnamefont{C.~O.} \bibnamefont{Yang}},
  \bibinfo{journal}{J.of Appl.Phys.} \textbf{\bibinfo{volume}{85}},
  \bibinfo{pages}{8446} (\bibinfo{year}{1999}).

\bibitem[{\citenamefont{Kichinosuke~Hirokawa and Oku}(1975)}]{Hirokawa:o}
\bibinfo{author}{\bibfnamefont{F.~H.} \bibnamefont{Kichinosuke~Hirokawa}}
  \bibnamefont{and} \bibinfo{author}{\bibfnamefont{M.}~\bibnamefont{Oku}},
  \bibinfo{journal}{J. Elec. Spec. Rel. Phenom.} \textbf{\bibinfo{volume}{6}},
  \bibinfo{pages}{333} (\bibinfo{year}{1975}).

\bibitem[{\citenamefont{Wu et~al.}(1994)\citenamefont{Wu, Chiou, and
  Hsieh}}]{wu:oxdef}
\bibinfo{author}{\bibfnamefont{W.-F.} \bibnamefont{Wu}},
  \bibinfo{author}{\bibfnamefont{B.-S.} \bibnamefont{Chiou}}, \bibnamefont{and}
  \bibinfo{author}{\bibfnamefont{S.-T.} \bibnamefont{Hsieh}},
  \bibinfo{journal}{semicond. Sci. Technol.} \textbf{\bibinfo{volume}{9}},
  \bibinfo{pages}{1242} (\bibinfo{year}{1994}).

\bibitem[{\citenamefont{Kim et~al.}(1999{\natexlab{b}})\citenamefont{Kim,
  Gilmore, Pique´, Horwitz, Mattoussi, Murata, Kafafi, and
  Chrisey}}]{kim:oxdef}
\bibinfo{author}{\bibfnamefont{H.}~\bibnamefont{Kim}},
  \bibinfo{author}{\bibfnamefont{C.~M.} \bibnamefont{Gilmore}},
  \bibinfo{author}{\bibfnamefont{A.}~\bibnamefont{Pique´}},
  \bibinfo{author}{\bibfnamefont{J.~S.} \bibnamefont{Horwitz}},
  \bibinfo{author}{\bibfnamefont{H.}~\bibnamefont{Mattoussi}},
  \bibinfo{author}{\bibfnamefont{H.}~\bibnamefont{Murata}},
  \bibinfo{author}{\bibfnamefont{Z.~H.} \bibnamefont{Kafafi}},
  \bibnamefont{and} \bibinfo{author}{\bibfnamefont{D.~B.}
  \bibnamefont{Chrisey}}, \bibinfo{journal}{J. of Appl. Phys.}
  \textbf{\bibinfo{volume}{86}}, \bibinfo{pages}{6451}
  (\bibinfo{year}{1999}{\natexlab{b}}).

\bibitem[{\citenamefont{Hollinger et~al.}(1975)\citenamefont{Hollinger, Jugnet,
  Pertosa, and Duc}}]{Hollinger:XPS}
\bibinfo{author}{\bibfnamefont{G.}~\bibnamefont{Hollinger}},
  \bibinfo{author}{\bibfnamefont{Y.}~\bibnamefont{Jugnet}},
  \bibinfo{author}{\bibfnamefont{P.}~\bibnamefont{Pertosa}}, \bibnamefont{and}
  \bibinfo{author}{\bibfnamefont{T.~M.} \bibnamefont{Duc}},
  \bibinfo{journal}{Chem. Phys. Lett.} \textbf{\bibinfo{volume}{36}},
  \bibinfo{pages}{441} (\bibinfo{year}{1975}).

\bibitem[{\citenamefont{Diplas et~al.}(2010)\citenamefont{Diplas, Lovvik,
  Nordmark, Kepaptsoglou, Graff, Ladam, Tyholdt, Walmsley, Gunnaes, Fagerberg
  et~al.}}]{diplas:ITO}
\bibinfo{author}{\bibfnamefont{S.}~\bibnamefont{Diplas}},
  \bibinfo{author}{\bibfnamefont{O.~M.} \bibnamefont{Lovvik}},
  \bibinfo{author}{\bibfnamefont{H.}~\bibnamefont{Nordmark}},
  \bibinfo{author}{\bibfnamefont{D.}~\bibnamefont{Kepaptsoglou}},
  \bibinfo{author}{\bibfnamefont{J.~M.} \bibnamefont{Graff}},
  \bibinfo{author}{\bibfnamefont{C.}~\bibnamefont{Ladam}},
  \bibinfo{author}{\bibfnamefont{F.}~\bibnamefont{Tyholdt}},
  \bibinfo{author}{\bibfnamefont{J.~C.} \bibnamefont{Walmsley}},
  \bibinfo{author}{\bibfnamefont{A.~E.} \bibnamefont{Gunnaes}},
  \bibinfo{author}{\bibfnamefont{R.}~\bibnamefont{Fagerberg}},
  \bibnamefont{et~al.}, \bibinfo{journal}{Surf. and Inter. Anal.}
  \textbf{\bibinfo{volume}{42}} (\bibinfo{year}{2010}).

\bibitem[{\citenamefont{Moshopoulou et~al.}(2006)\citenamefont{Moshopoulou,
  Ibberson, Sarrao, Thompson, and Fisk}}]{In:ref}
\bibinfo{author}{\bibfnamefont{E.~G.} \bibnamefont{Moshopoulou}},
  \bibinfo{author}{\bibfnamefont{R.~M.} \bibnamefont{Ibberson}},
  \bibinfo{author}{\bibfnamefont{J.~L.} \bibnamefont{Sarrao}},
  \bibinfo{author}{\bibfnamefont{J.~D.} \bibnamefont{Thompson}},
  \bibnamefont{and} \bibinfo{author}{\bibfnamefont{Z.}~\bibnamefont{Fisk}},
  \bibinfo{journal}{Acta Crystallographica B} \textbf{\bibinfo{volume}{39}},
  \bibinfo{pages}{173} (\bibinfo{year}{2006}).

\end{thebibliography}

\end{document}